\documentclass[twocolumn,aps,superscriptaddress,preprintnumbers,amsmath,amssymb]{revtex4}

\usepackage{graphicx}
\usepackage{dcolumn}
\usepackage{bm}
\usepackage{color}

\begin{document}


\title{Avoided quantum criticality and cluster-glass formation in itinerant ferromagnet Sr$_{\bm{1-x}}$(La$_{\bm{0.5}}$K$_{\bm{0.5}}$)$_{\bm{x}}$RuO$_{\bm{3}}$}
\author{Ryoya~Iwahara}
\author{Ryoma~Sugawara}
\author{Rahmanto}
\author{Yutoku~Honma}
\author{Kensuke~Matsuoka}
\affiliation{Faculty of Science, Ibaraki University, Mito, Ibaraki 310-8512, Japan\\
and Institute of Quantum Beam Science, Ibaraki University, Mito, Ibaraki 310-8512, Japan}
\author{Akira~Matsuo}
\author{Koichi~Kindo}
\affiliation{Institute for Solid State Physics, The University of Tokyo, Kashiwa, Chiba 277-8581, Japan}
\author{Kenichi~Tenya}
\affiliation{Faculty of Education, Shinshu University, Nagano 380-8544, Japan}
\author{Makoto~Yokoyama}
\email[]{makoto.yokoyama.sci@vc.ibaraki.ac.jp}
\affiliation{Faculty of Science, Ibaraki University, Mito, Ibaraki 310-8512, Japan\\
and Institute of Quantum Beam Science, Ibaraki University, Mito, Ibaraki 310-8512, Japan}

\date{\today}
             
\begin{abstract}
We demonstrate that the cluster-glass state emerges as ferromagnetic quantum criticality is avoided in itinerant ferromagnet Sr$_{1-x}$(La$_{0.5}$K$_{0.5}$)$_x$RuO$_3$. In this compound, the ferromagnetic order is suppressed by increasing $x$, and then disappears at the critical concentration: $x=0.5$. In this $x$ range, the present study reveals that no prominent feature is ascribed to the quantum critical fluctuations in specific heat. Instead, ac magnetic susceptibility exhibits a broad peak due to spontaneous spin freezing, and the peak temperature depends significantly on the frequency of the applied ac magnetic field. Furthermore, specific heat is enhanced within a wide temperature range, whereas specific heat shows no salient anomaly associated with spin freezing. These features are characteristics of the formation of cluster-glass; in particular, the observed frequency variations in ac magnetic susceptibility are well described by the Vogel-Fulcher law.  We compare the features concerning the suppression of the ferromagnetic order in this doped compound with those in isostructural Ca- and La-doped SrRuO$_3$, and suggest that a local correlated disorder effect and the very small coherence of itinerant Ru 4$d$ electrons are responsible for the cluster-glass formation instead of the quantum phase transition in Sr$_{1-x}$(La$_{0.5}$K$_{0.5}$)$_x$RuO$_3$.
\end{abstract}

\maketitle

\section{Introduction}
Understanding the role of spin correlations in the anomalous electronic state is one of the most challenging subjects in strongly correlated electron physics related to the metal-insulator (MI) transition. The distorted perovskite compound SrRuO$_3$ [GdFeO$_3$-type orthorhombic crystal structure, the upper inset in Fig.\ 1(a)] shows the ferromagnetic (FM) order below $T_{\rm C}=160\ {\rm K}$ \cite{rf:Callaghan66,rf:Kanbayasi76}, whose order parameter is considered to originate from the itinerant Ru 4$d$ electrons from early photoemission and band-calculation studies \cite{rf:Allen96,rf:Fujioka97,rf:Okamoto99}. However, this compound concomitantly involves so-called bad metallic characteristics, corresponding to the absence of suppression in electrical resistivity and a very small mean-free path comparable to the lattice constants at high temperatures \cite{rf:Klein96,rf:Laad2001,rf:Gunnarsson2003,rf:Cao2004}. Optical conductivity exhibits anomalous charge dynamics which are different from those expected from conventional Fermi-liquid behaviors at low temperatures \cite{rf:Kostic98}. Furthermore, angle-resolved photoemission investigations indicate that the localized spin state of Ru 4$d$ electrons emerges in the paramagnetic region \cite{rf:Shai2013}. The Rhodes-Wohlfarth parameter, which is determined by the ratio of a paramagnetic moment and an ordered FM moment, is estimated to be about 1.3 in SrRuO$_3$, which does not coincide with the localized spin limit ($=1$) or the itinerant spin limit ($\sim 3$) \cite{rf:Fukunaga94}. These features are suggested to be a signature of the itinerant-localized dual nature in Ru 4$d$ electrons which is enhanced due to the instability of the MI transition.

The MI transition is induced in mixed compounds of SrRuO$_3$, such as SrRu$_{1-x}$Mn$_x$O$_3$ \cite{rf:Sahu2002,rf:Cao2005,rf:Yokoyama2005,rf:Zhang2007,rf:Kolesnik2008}, SrRu$_{1-x}$Rh$_x$O$_3$ \cite{rf:Yamaura2004}, and SrRu$_{1-x}$Mg$_x$O$_3$ \cite{rf:Crandles2002}. In contrast, the substitution of Ca or La for Sr in SrRuO$_3$ does not induce the MI transition, but yields anomalous paramagnetic ground states through suppression of the FM phase. It was revealed that in Sr$_{1-x}$Ca$_x$RuO$_3$, doping Ca into SrRuO$_3$ monotonically suppresses the FM order, and then yields a paramagnetic state above a critical concentration: $x \sim 0.7-0.8$ \cite{rf:Cao97}. In this $x$ range, nuclear magnetic resonance (NMR) and thermodynamic investigations suggest an evolution of FM quantum-critical fluctuations originating from the itinerant spins \cite{rf:Yoshimura97,rf:Kiyama98,rf:Huang2015}. However, muon spin rotation ($\mu$SR) and magneto-optical experiments point out that such spin fluctuations tend to be weakened or smeared by a spontaneous phase separation between the FM and paramagnetic states in real space \cite{rf:Uemura2007,rf:Demko2012}. In Sr$_{1-x}$La$_x$RuO$_3$, the FM ordered state is rapidly suppressed by La doping \cite{rf:Bouchard72,rf:Nakatsugawa2002,rf:Xu2018}, and is then replaced by a cluster-glass state for $0.3 \le x \le 0.5$, shown by the experimental result that ac magnetic susceptibility shows frequency dependence near the cluster-glass freezing temperature \cite{rf:Kawasaki2014}. The development of inhomogeneous FM clusters was also confirmed with $\mu$SR measurement \cite{rf:Kawasaki2016-1}. Furthermore, photoemisson experiments revealed that most of the coherent component of Rh 4$d$ electrons remains at the Fermi level in the photoemission spectra for $0.3 \le x \le 0.5$, although the spectral weight transfers from the coherent to incoherent parts with increasing $x$ \cite{rf:Kawasaki2016-2}. It is expected that these anomalous features in Ca- and La-doped SrRuO$_3$ are intimately coupled with the itinerant-localized duality involved in Ru 4$d$ electrons.   

To gain further insight into the anomalous electronic and spin states enhanced around the FM critical region, we investigated the other mixed compound Sr$_{1-x}$(La$_{0.5}$K$_{0.5}$)$_x$RuO$_3$ by performing magnetic and thermal experiments. A previous report on this compound indicated that the overall feature concerning suppression of the FM order is similar to those observed in Ca- and La-doped SrRuO$_3$, and the critcal $x$ value for the disappearance of the FM order is $\sim 0.5$ \cite{rf:Shuba2006}. However, the nature of the electronic and spin states around the critical $x$ range remains unclear. As for the case of Ca- and La-doped SrRuO$_3$, the striking difference between the dopants is considered to be their nominal valence states; no carrier-doping effect is expected in Ca-doped SrRuO$_3$ because the Sr$^{2+}$ ion is substituted by the isovalent Ca$^{2+}$ ion, whereas the La$^{3+}$ ion is considered to behave as an electron dopant as well as an impurity in La-doped SrRuO$_3$. We expect that our investigation of Sr$_{1-x}$(La$_{0.5}$K$_{0.5}$)$_x$RuO$_3$ will provide clues for understanding the role of this discrepancy in the evolution of anomalous electronic and spin states, because no carrier-doping effect is expected with equal amounts of La$^{3+}$ and K$^{+}$ ion doping. In this paper, we demonstrate the emergence of the cluster-glass state for $0.4 \le x \le 0.47$, while Ru 4$d$ electrons simultaneously involve itinerant characteristics. Then we discuss the similarities and differences among the doped compounds to clarify the origin of anomalous electronic and spin states at the FM critical region.     

\section{Experiment Details}
Polycrystals of Sr$_{1-x}$(La$_{0.5}$K$_{0.5}$)$_x$RuO$_3$ with $x \le 0.5$ were synthesized with a conventional solid-state reaction method with starting materials SrCO$_3$, La(OH)$_3$, K$_2$CO$_3$, and RuO$_2$. To make the samples, the SrCO$_3$, La(OH)$_3$, and K$_2$CO$_3$ powders with stoichiometric compositions were initially mixed and calcined at 800 $^{\circ}$C, and the products were then mixed with the RuO$_2$ powder in high-purity ethanol. After the ethanol was removed at 100 $^{\circ}$C, the mixtures were shaped into pellets, and sintered at 1250 $^{\circ}$C for 20 h in ambient atmosphere. This sintering process was iterated 2 to 3 times to ensure homogeneous synthesis. The synthesis procedure for the pure SrRuO$_3$ sample is described elsewhere \cite{rf:Kawasaki2014}. The x-ray diffraction measurements confirmed that all the samples had the GdFeO$_3$-type orthorhombic crystal structure and no extrinsic phase within experimental accuracy. The lattice parameters were consistent with those in a previous report \cite{rf:Shuba2006}.

\begin{figure}[tbp] 
\begin{center}
\includegraphics[viewport=0 0 488 608,keepaspectratio,width=0.42\textwidth]{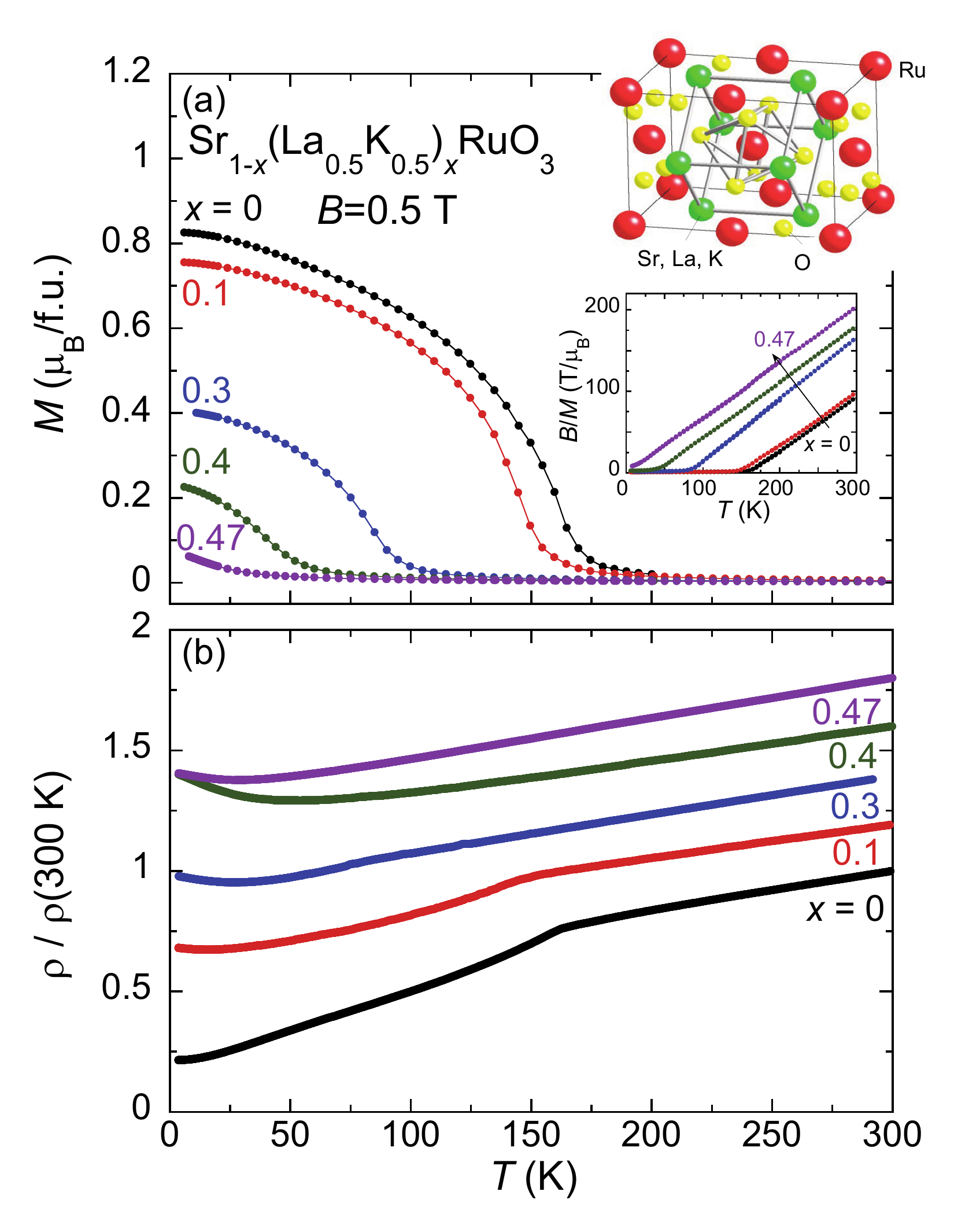}
\end{center}
\vspace{-12pt}
  \caption{
Temperature variations in (a) dc magnetization $M$ obtained under magnetic field of $B=0.5\ {\rm T}$ with the field-cooling condition and (b) electrical resistivity $\rho$ normalized by the magnitude at 300 K for Sr$_{1-x}$(La$_{0.5}$K$_{0.5}$)$_x$RuO$_3$. The $\rho/\rho(300\ {\rm K})$ data for $x \ge 0.1$ are vertically shifted in 0.2 steps for clarity. The crystal structure and inverse susceptibility $B/M$ of Sr$_{1-x}$(La$_{0.5}$K$_{0.5}$)$_x$RuO$_3$ are shown in the upper and lower insets, respectively, in (a).
}
\end{figure}
Displayed in Fig.\ 1(a) are the temperature variations in dc magnetization $M(T)$ for Sr$_{1-x}$(La$_{0.5}$K$_{0.5}$)$_x$RuO$_3$, measured using a commercial SQUID magnetometer (MPMS, Quantum Design). The overall features of the $M(T)$ data are consistent with those reported in previous work \cite{rf:Shuba2006}; the Curie temperature and the magnitude of spontaneous magnetization are reduced with increasing $x$, and then become nearly zero at $x\sim 0.5$. In addition, the effective moments estimated from inverse magnetic susceptibility [the lower inset in Fig.\ 1(a)] in the paramagnetic region are 2.6(2) $\mu_{B}/{\rm f.u.}$ for all of the $x$ range investigated, which roughly coincide with the calculated value (2.8 $\mu_{B}$) for the low-spin state of the Ru$^{4+}$ ion. 

We also checked the temperature dependence in electrical resistivity $\rho(T)$ for our Sr$_{1-x}$(La$_{0.5}$K$_{0.5}$)$_x$RuO$_3$ samples [Fig.\ 1(b)]. The magnitude of $\rho$ at 300 K ranges from 0.8 to 2.0 m$\Omega$ cm for all of the sample investigated, and this discrepancy may be caused by the sintered samples. For $x=0$, a kink is observed in $\rho(T)$ at $T_{\rm C}=160\ {\rm K}$, and this feature becomes unclear along with the reduction of $T_{\rm C}$ as $x$ is increased. At the same time, a slight upturn emerges in $\rho(T)$ at low temperatures for $x\ge 0.3$. This upturn feature was also observed in the previous investigations for (La,K)-, La-, and Ca-doped SrRuO$_3$ \cite{rf:Kawasaki2014,rf:Shuba2006,rf:Ahn99}, but its origin is unclear at present. Despite the weak upturn feature, it is considered that the $\rho(T)$ data for all of the $x$ range investigated have metallic characteristics. These trends are consistent with the previous reports \cite{rf:Shuba2006}.

The ac magnetic susceptibility measurements were performed between 4 and 220 K with the standard Hartshorn-bridge method, in which the magnitude and the frequency of the applied ac field were selected to be 0.1 mT and 12--1020 Hz, respectively. The specific heat measurements were carried out down to 1.1 K with the thermal relaxation method, in which well-defined thermal-relaxation curves were obtained for all the measurements by carefully setting the thermal contact between the plate-shaped sintered samples and the heat-capacity chip of the equipment.

\section{Results}
\subsection{Magnetic properties}
\begin{figure}[tbp] 
\begin{center}
\includegraphics[viewport=0 0 450 528,keepaspectratio,width=0.45\textwidth]{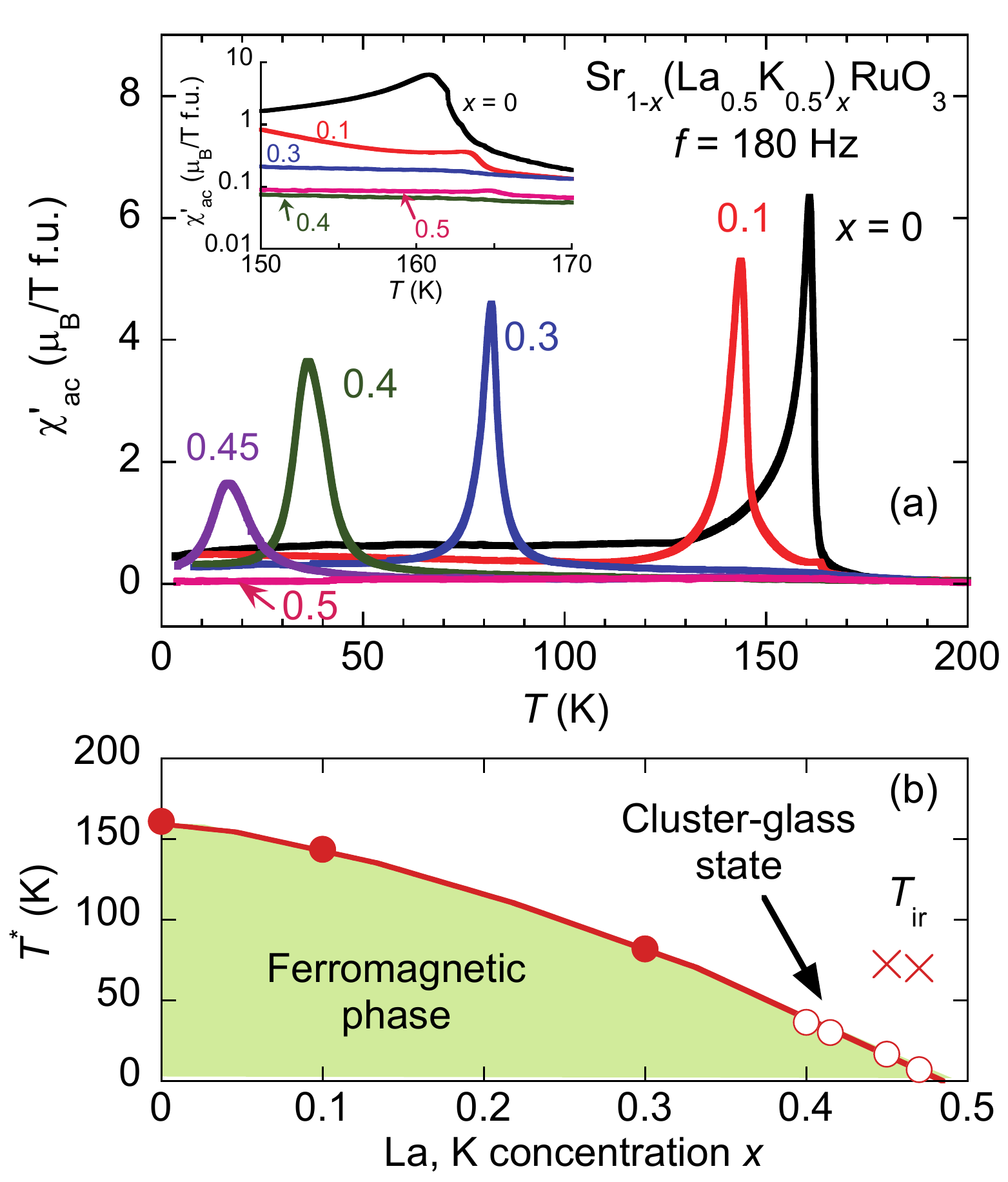}
\end{center}
\vspace{-12pt}
  \caption{
(a) Temperature variations in the real-part component in ac magnetic susceptibility $\chi_{\rm ac}'$ at 180 Hz, and (b) $x$ variations of peak temperature $T^*$ estimated from the $\chi_{\rm ac}'(T)$ data for Sr$_{1-x}$(La$_{0.5}$K$_{0.5}$)$_x$RuO$_3$. The inset in (a) shows the increase in the $\chi_{\rm ac}'$ data around 160 K, in which a logarithmic scale is used for the vertical axis. In (b), the open and cross symbols indicate the freezing temperature of the cluster-glass state and the characteristic temperature $T_{\rm ir}$ below which the $M(T)$ curve for the zero-field-cooling condition deviates from that for the field-cooling condition, respectively.
}
\end{figure}
Figure 2(a) shows the temperature dependence of the real-part component in ac magnetic susceptibility $\chi_{\rm ac}'(T)$ with frequency $f$ of 180 Hz for Sr$_{1-x}$(La$_{0.5}$K$_{0.5}$)$_x$RuO$_3$. A clear divergence associated with the FM transition is observed at 161 K in  $\chi_{\rm ac}'(T)$ for pure SrRuO$_3$. The doping of La and K into SrRuO$_3$ rapidly reduces peak temperature $T^*$ in $\chi_{\rm ac}'(T)$ down to 144 K ($x=0.1$) and 82 K ($x=0.3$), while the diverging feature remains for $x \le 0.3$.  As $x$ approaches the FM critical concentration ($\sim 0.5$), however, the peak in $\chi_{\rm ac}'(T)$ becomes small and broad; the peak height and width for $x=0.45$ ($T^*=16.6\ {\rm K}$) are about one fourth of and threefold those for $x=0$, respectively. Finally, the peak is not detected in $\chi_{\rm ac}'(T)$ for $x=0.5$ within experimental accuracy, at least in the temperature range of $T \ge 4\ {\rm K}$. The diagram of $x$ versus $T^*$ [Fig.\ 2(b)] is consistent with that derived from dc magnetization \cite{rf:Shuba2006}. Note that a tiny peak at $\sim 163$ K is observed in the $\chi_{\rm ac}'(T)$ data for $x=0.1$ [the inset in Fig.\ 2(a)]. This peak is likely caused by a fragmentary phase of pure SrRuO$_3$ because the magnitude of this peak component is only 1.6\% of that of pure SrRuO$_3$, but the sharpness of the peak is comparable. For $x=0.45$ [not shown in the inset in Fig.\ 2(a)] and 0.5, $\chi_{\rm ac}'(T)$ exhibits an extremely small hump at $\sim 165\ {\rm\ K}$, whose magnitude is about 0.1\% of the peak height of pure SrRuO$_3$. In contrast, no such a peak is in $\chi_{\rm ac}'(T)$ for the other compositions.

\begin{figure}[tbp] 
\begin{center}
\includegraphics[viewport=0 0 456 714,keepaspectratio,width=0.42\textwidth]{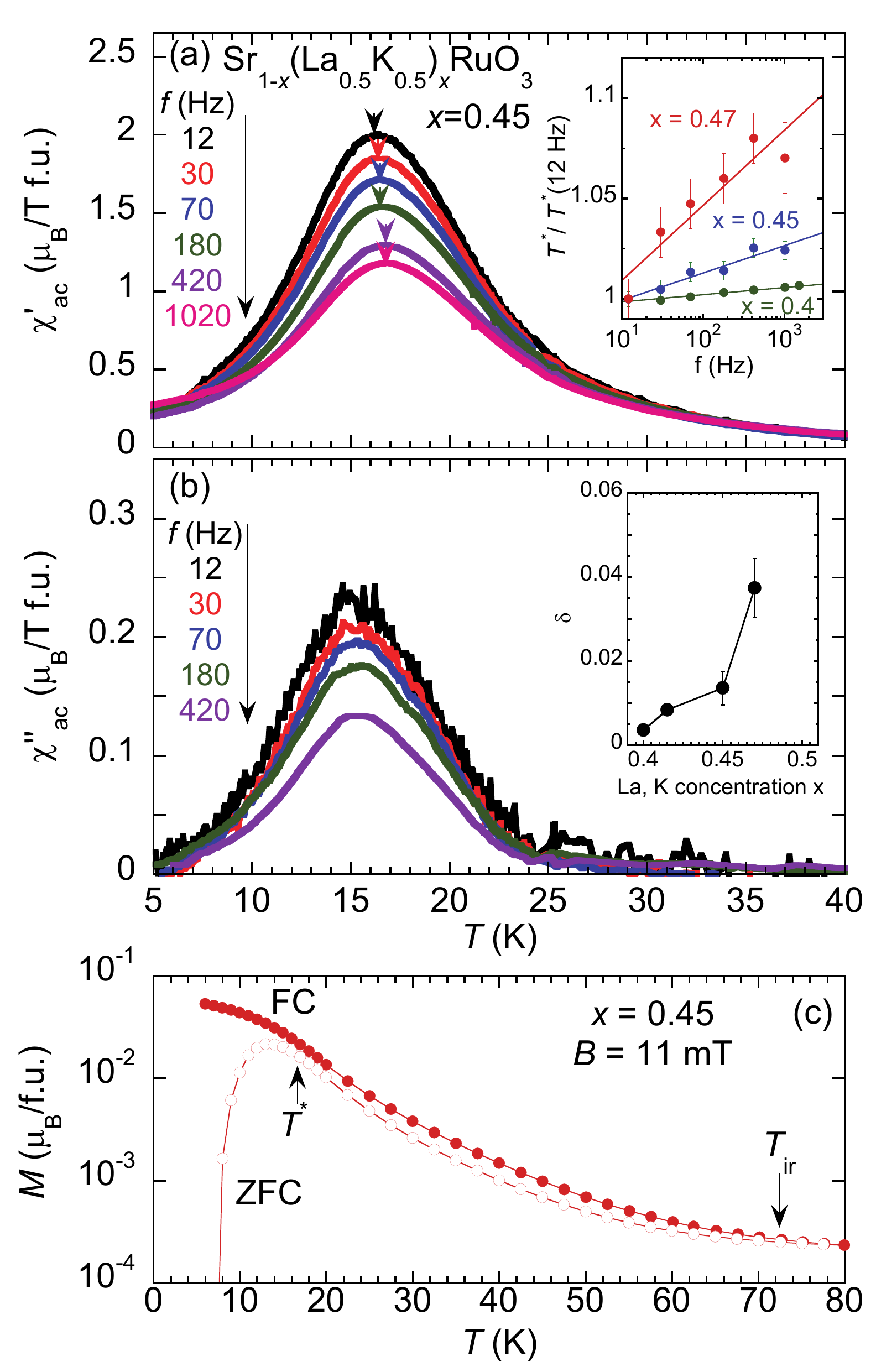}
\end{center}
\vspace{-12pt}
  \caption{
Temperature variations in (a) real-part ac susceptibility $\chi_{\rm ac}'$ and (b) imaginary-part ac susceptibility $\chi_{\rm ac}''$ for Sr$_{0.55}$(La$_{0.5}$K$_{0.5}$)$_{0.45}$RuO$_3$, obtained under the ac magnetic field with various frequencies. The arrows in (a) indicate peak temperature $T^*$ of  $\chi_{\rm ac}'$ for different frequencies. The inset of (a) shows the frequency dependencies of $T^*$ for $x=0.4$, 0.45, and 0.47, and the inset of (b) is the $x$ dependence of the initial frequency shift $\delta$ for $0.4 \le x \le 0.47$. (c) Temperature variations in dc magnetization for Sr$_{0.55}$(La$_{0.5}$K$_{0.5}$)$_{0.45}$RuO$_3$, obtained under very weak magnetic field of $B=11\ {\rm mT}$ with the field-cooling (FC) and zero-field-cooling (ZFC) conditions.
}
\end{figure}
The suppression and broadening of the peak in $\chi_{\rm ac}'(T)$ for $x\sim 0.45$ indicate an increase in the disorder and dynamical effects on the spin arrangement around the FM critical region. To verify those effects, we measured the frequency dependence of the peak in ac magnetic susceptibility. Figure 3(a) shows the increase in the $\chi_{\rm ac}'(T)$ data around $T^*$ for $x=0.45$, obtained under the ac magnetic field with different frequencies. A clear peak shift toward high temperatures and a reduction in the peak height occur with the increasing frequency. In addition, the small peak evolves in the imaginary-part component of ac magnetic susceptibility $\chi_{\rm ac}''(T)$ at $\sim T^*$ [Fig.\ 3(b)], indicating that an energy dissipation process is involved in the spin arrangement at $\sim T^*$. All these features strongly suggest that glass-like spin freezing occurs at $\sim T^*$. A similar frequency dependence is observed in $\chi_{\rm ac}'(T)$ for $0.4 \le x \le 0.47$, although this feature is not detected for $x \le 0.3$ within experimental accuracy. Furthermore, the rate of the frequency shift in $T^*$ increases with the increasing $x$ [the inset in Fig.\ 3(a)]. Note that the frequency range of this investigation ($\le1020\ {\rm Hz}$) is much smaller than the megahertz and gigahertz ranges in which magnetic responses are expected to show the frequency dependence in usual FM compounds \cite{rf:Mydosh93}.

In general, the characteristics of glass-like freezing of the spins can be inferred from the value of the initial frequency shift (the Mydosh parameter) defined as $\delta=\Delta T^*/(T^*\Delta \log_{10}f)$ \cite{rf:Mydosh93}. In the inset in Fig.\ 3(b), we plot the $\delta$ value estimated from the $\chi_{\rm ac}'(T)$ data for $0.4 \le x \le 0.47$. $\delta$ increases with increasing $x$, ranging from 0.0036(2) ($x=0.4$) to 0.037(7) ($x=0.47$). Although the $\delta$ values for $x=0.4$ and 0.415 are comparable to those estimated for the canonical spin-glass system of CuMn ($\delta \sim 0.005$), the values are increased to the order of 0.01 for $x \ge 0.45$, which covers the $\delta$ range expected for the cluster-glass state ($\delta \sim$ 0.01--0.1), realized as an ensemble of interacting spin nano clusters \cite{rf:Mydosh93}. This suggests that the increase in $\delta$ is ascribed to the variation in the spin arrangement from the FM long-range order to the FM cluster-glass formation as $x$ approaches the FM critical concentration of $x\sim 0.5$.     

The emergence of the FM cluster-glass state can be also traced in the $M(T)$ curves. Figure 3(c) shows $M(T)$ for $x=0.45$ obtained under very weak magnetic field of $B=11\ {\rm mT}$ with the field-cooling (FC) and zero-field-cooling (ZFC) conditions. It is found that $M(T)$ for the ZFC condition deviates from that for the FC condition below $T_{\rm ir}\sim 72.5\ {\rm K}$, and then exhibits a peak at 13.5 K, followed by the large reduction of $M(T)$ with further decreasing temperature. The discrepancy between the peak temperature and $T^*\ (=16.6\ {\rm K})$ is caused by the magnitude of the applied magnetic field, because the peak temperature is found to approach to $T^*$ with further decreasing $B$. In contrast, $M(T)$ for the FC condition continuously increases with decreasing temperature. These behaviors in $M(T)$ are the characteristics of the formation of the FM cluster-glass state; it is expected that the FM clusters develop and start interacting (or freezing) below $\sim T_{\rm ir}$, and the majority of the FM clusters then freeze below $\sim T^*$, yielding the very small and large hysteresis below $T_{\rm ir}$ and $T^*$, respectively, in $M(T)$ between the FC and ZFC conditions. These features are also observed in the other cluster-glass systems \cite{rf:Tian2016,rf:Yamamoto2016}.

\begin{figure}[tbp] 
\begin{center}
\includegraphics[viewport=0 0 478 388,keepaspectratio,width=0.46\textwidth]{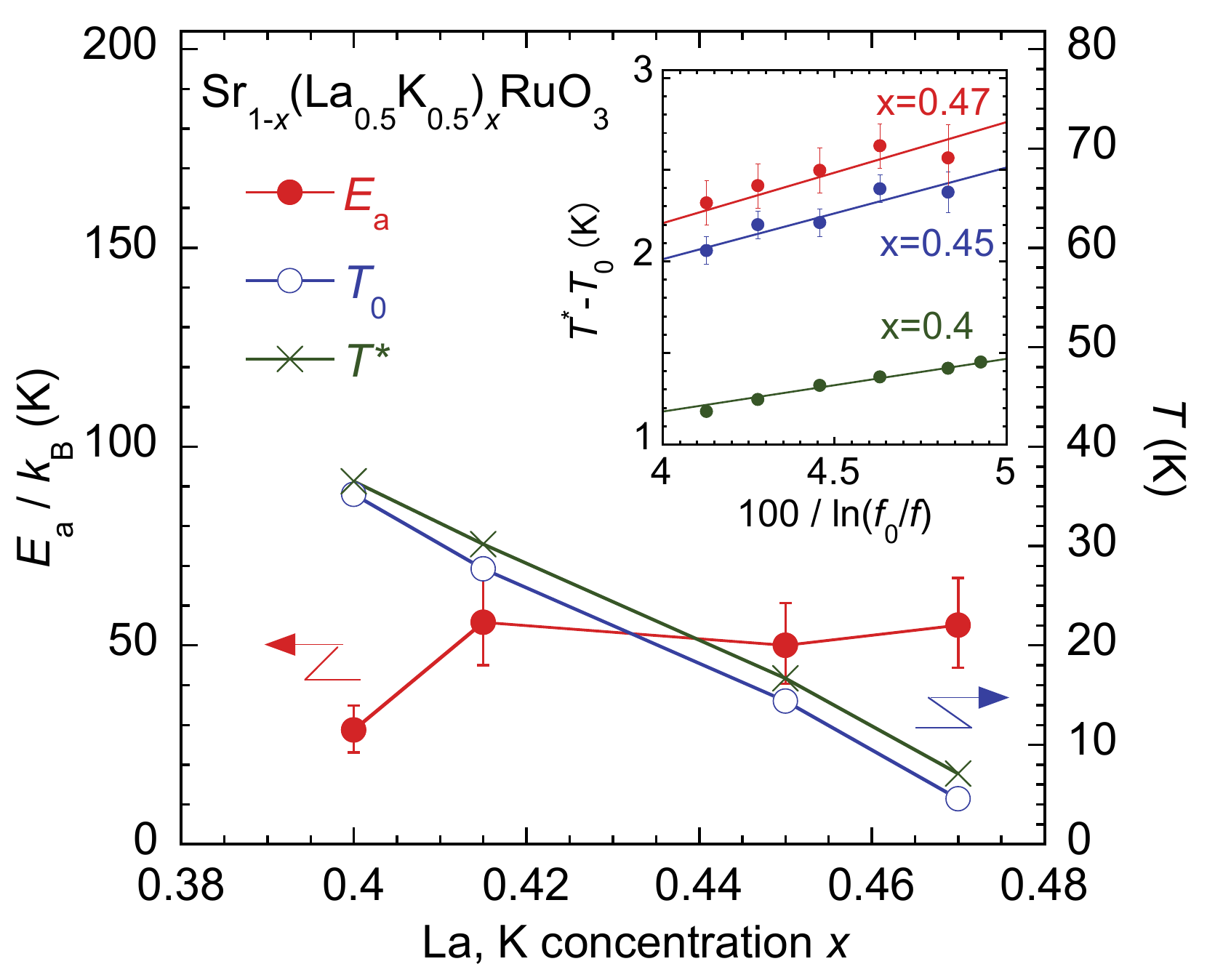}
\end{center}
\vspace{-12pt}
  \caption{
Activation energy $E_{\rm a}$ and Vogel-Fulcher temperature $T_0$ around the FM critical $x$ range ($0.4 \le x \le 0.47$) for Sr$_{1-x}$(La$_{0.5}$K$_{0.5}$)$_{x}$RuO$_3$, plotted as a function of the La and K concentration $x$. In this plot, $f_0$ is assumed to be 10$^{12}$ Hz. The error bar of $E_{\rm a}$ originates from the $f_0$ range from 10$^{11}$ to 10$^{13}$ Hz. The $T^*$ data at 180 Hz are also plotted for comparison. The inset shows the $T^*$ data for $x=0.4$, 0.45, and 0.47 plotted as a function of $100/\ln (f_0/f)$, in which the $T_0$ value obtained with Vogel-Fulcher analysis for each $x$ is subtracted from $T^*$ for clarity. 
}
\end{figure}
To clarify the nature of the cluster-glass state for $0.4 \le x \le 0.47$, we attempt phenomenological Vogel-Fulcher analysis for the $T^*$ data obtained from $\chi_{\rm ac}'(T)$. In the Vogel-Fulcher law, the relation between $f$ and $T^*$ is described by
\begin{eqnarray}
f=f_0\exp\left[-\frac{E_{\rm a}}{k_{\rm B}(T^*-T_0)}\right],
\end{eqnarray}
where $f_0$, $E_{\rm a}$, and $T_0$ are the characteristic frequency of the spin clusters, the activation energy, and the Vogel-Fulcher temperature, respectively. In general, the $f_0$ value is known to fall within the range between 10$^{11}$ and 10$^{13}$ Hz for typical spin and cluster glass systems \cite{rf:Dho2002,rf:Cardoso2003}, and therefore, we tentatively assume that $f_0$ is 10$^{12}$ Hz. It is convenient to rewrite Eq.\ (1) for analyzing the frequency dependence of $T^*$ as follows:
\begin{eqnarray}
T^*=T_0+\frac{E_{\rm a}}{k_{\rm B}}\left[\ln\left(\frac{f_0}{f} \right)\right]^{-1}.
\end{eqnarray}
The present $T^*$ data are roughly in proportion to $1/\ln(1/f)$, at least within the investigated $f$ range (the inset in Fig.\ 4). Figure 4 shows the $E_{\rm a}$ and $T_0$ values for $0.4 \le x \le 0.47$, obtained with the best fit of the $T^*$ data using Eq.\ (2). In this figure, the error bar of $E_{\rm a}$ originates from the difference in the assumed $f_0$ values between 10$^{11}$ and 10$^{13}$ Hz. The near coincidence between $T_0$ and $T^*$ and the relation of $T_0 < T^*$ confirm that the peak in $\chi_{\rm ac}'(T)$ is attributed to freezing of the spin clusters because $T_0$ is considered to be related to the strength of the inter-cluster interactions. It is also found that the $E_{\rm a}/k_{\rm B}$ values ($\sim$ 30--50 K) are comparable to the magnitude of $T_{\rm ir}$ as well as $T^*$ for $x=0.4$, at which the crossover from the FM long-range order to the cluster-glass state occurs. Furthermore, $E_{\rm a}$ is roughly independent of $x$ for $0.415 \le x \le 0.47$. These features indicate that the energy barrier of spin cluster flipping does not depend much on $x$, and the suppression of $T^*$ in this $x$ range is mainly attributed to the reduction in inter-cluster interactions rather than the shrinkage of each spin cluster. This trend is also seen in Sr$_{1-x}$La$_x$RuO$_3$ for $0.3 \le x \le 0.5$ \cite{rf:Kawasaki2014}. Note that the Arrhenius equation, given by putting $T_0=0$ in Eq.\ (1), cannot be applied to the $T^*$ data for Sr$_{1-x}$(La$_{0.5}$K$_{0.5}$)$_{x}$RuO$_3$ because the best fit using the Arrhenius law leads to inappropriate fitting parameters of $f_0 > 10^{29}\ {\rm Hz}$ and $E_{\rm a}/k_{\rm B}> 450\ {\rm K}$ for $0.4 \le x \le 0.47$. We expect that the peak in $\chi_{\rm ac}'(T)$ roughly follows the Arrhenius law for $x \ge 0.5$ if the $T_0=0$ condition is realized. However, the peak due to reduced spin cluster flipping is not observed in $\chi_{\rm ac}'(T)$ for $x=0.5$ in the present temperature range ($T \ge 4\ {\rm K}$).

\subsection{Thermal properties}
\begin{figure}[tbp] 
\begin{center}
\includegraphics[viewport=0 0 348 410,keepaspectratio,width=0.42\textwidth]{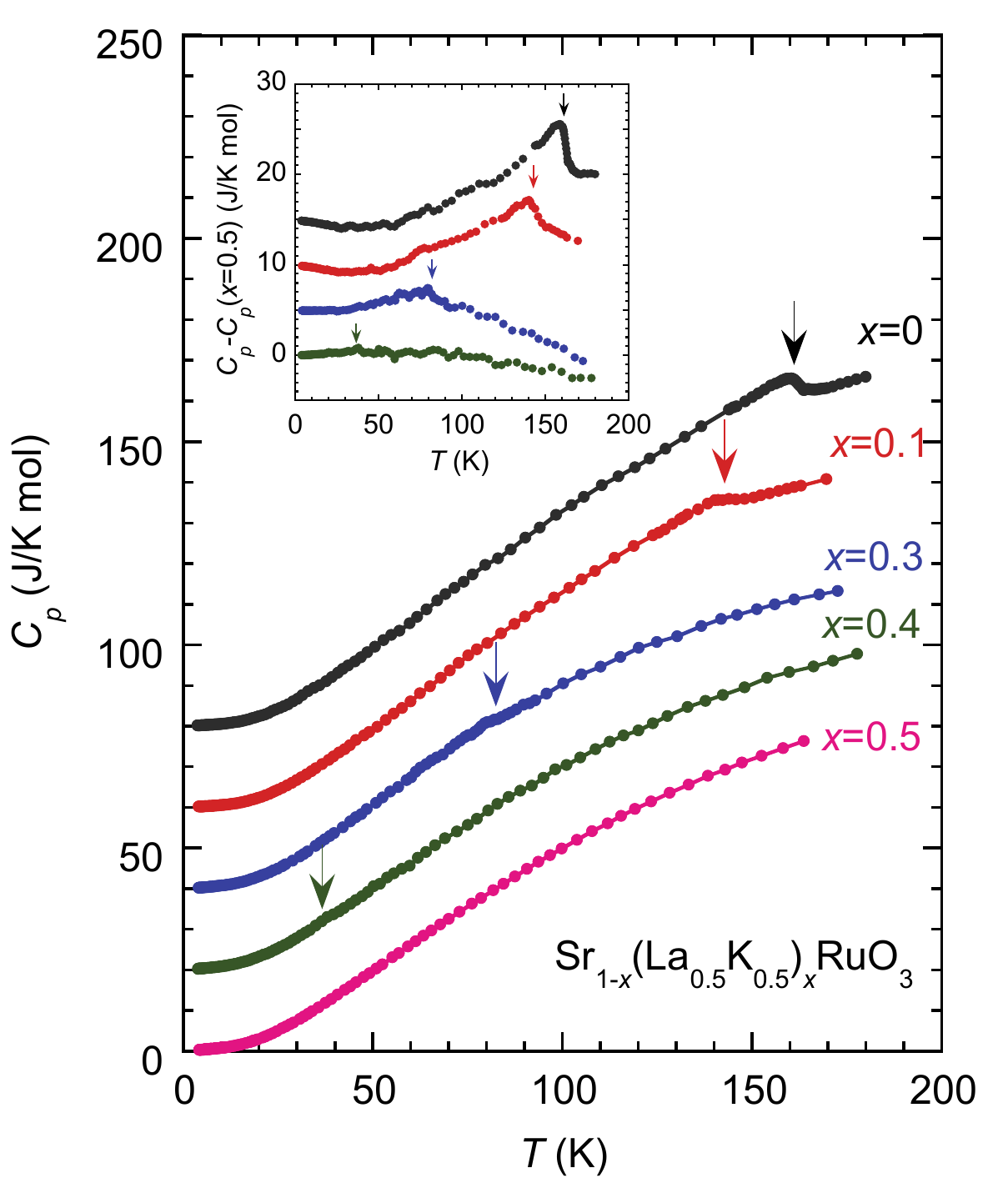}
\end{center}
\vspace{-12pt}
  \caption{
Temperature variations in specific heat $C_p$ for Sr$_{1-x}$(La$_{0.5}$K$_{0.5}$)$_{x}$RuO$_3$. Note that the $C_p$ data for $x \le 0.4$ are vertically shifted in 20 J/K mol steps for clarity. The inset shows the differences in the $C_p$ data from those for $x=0.5$, $C_p-C_p(x=0.5)$, in which the data for $x \le 0.3$ are vertically shifted in 5 J/K mol steps. The arrows indicate $T^*$ estimated from ac magnetic susceptibility.
}
\end{figure}
Figure 5 shows the temperature variations in specific heat $C_p(T)$ for Sr$_{1-x}$(La$_{0.5}$K$_{0.5}$)$_{x}$RuO$_3$. A clear jump associated with the FM transition occurs at $T^*$ (=161 K) in $C_p(T)$ for $x=0$. The jump in $C_p(T)$ is reduced and becomes broad with increasing $x$. The FM transition is still recognized in $C_p(T)$ as a weak kink at $T^*=82\ {\rm K}$ for $x=0.3$, but no clear anomaly due to phase transition is observed at $T^*$ in $C_p(T)$ for $x \ge 0.4$ within the experimental resolution. This feature can be more clearly seen in the inset in Fig.\ 5, where the differences in the $C_p$ data from those for the compound showing no magnetic order ($x=0.5$), $C_p-C_p(x=0.5)$, are plotted. The suppression of the jump in $C_p(T)$ seems consistent with the trends observed in the $M(T)$ and $\chi'_{\rm ac}(T)$ data. Namely, doping La and K reduces the magnitude of spontaneous magnetization in $M(T)$ and broadens the peak associated with the FM transition at $T^*$ in $\chi'_{\rm ac}(T)$. In addition, the frequency dependence of the peak at $T^*$ in $\chi'_{\rm ac}(T)$ indicates that the FM transition changes into random freezing of spin clusters above $x\sim 0.4$. These features in $M(T)$ and $\chi'_{\rm ac}(T)$ suggest that the entropy change associated with spin freezing is small at $T^*$, leading to the reduction in the anomaly at $T^*$ in $C_p(T)$. This trend in $C_p(T)$ is further discussed later.

\begin{figure}[tbp] 
\begin{center}
\includegraphics[viewport=0 0 470 605,keepaspectratio,width=0.42\textwidth]{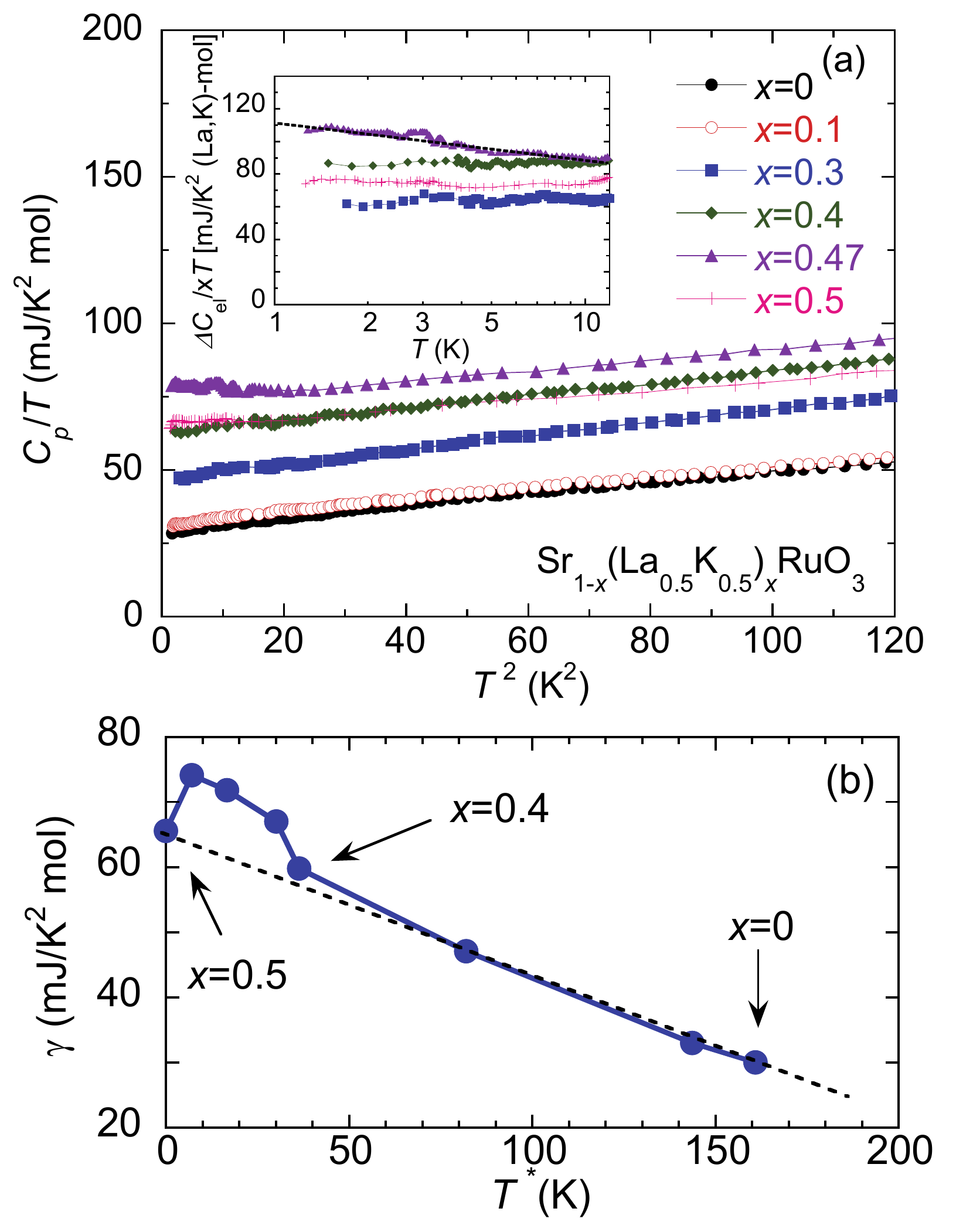}
\end{center}
\vspace{-12pt}
  \caption{
(a) Low-temperature specific heat divided by temperature $C_p/T$ plotted as a function of $T^2$ and (b) the relation between the electronic specific heat coefficient $\gamma$ and $T^*$, obtained using $x$ as an implicit parameter for Sr$_{1-x}$(La$_{0.5}$K$_{0.5}$)$_{x}$RuO$_3$. The inset in (a) shows the temperature dependence in the doping contribution to the electronic specific heat, $\Delta C_{\rm el}/xT \equiv[(C_{\rm el}-C_{\rm el}(x=0))/xT]$, in which a logarithmic scale is used for the horizontal axis. The dashed lines in the inset of (a) and (b) are visual guides.
}
\end{figure}
In Fig.\ 6(a), low-temperature specific heat divided by temperature $C_p/T$ is plotted as a function of $T^2$. The $C_p/T$ curves for the entire La and K concentration range of $x\le 0.5$ are roughly in proportion to $T^2$, and show no salient feature ascribed to quantum critical fluctuations, such as strongly diverging behavior for $T \to 0$. For $x=0.47$, the $C_p/T$ curve does not involve a pronounced anomaly due to the freezing of the spin clusters at $(T^*)^2=50.4\ {\rm K^2}$ ($T^*=7.1\ {\rm K}$), but it seems to exhibit a slight deviation from the $T^2$ function below $T^2\sim 20\ {\rm K}^2$ ($T\sim 4.5 \ {\rm K}$). Therefore, we simply assume a relation of $C_p/T=\gamma + \beta T^2$, and fit the $C_p/T$ data between 25 and 100 K$^2$ (5 and 10 K) using this function. Then, we estimate the electronic specific heat using a relation of $C_{\rm el}\equiv C_p-\beta T^3$. It is found that for $x=0.47$, the doping contribution to the electronic specific heat, $\Delta C_{\rm el}/xT \equiv[(C_{\rm el}-C_{\rm el}(x=0))/xT]$, slightly increases with decreasing temperature, roughly in proportion to $-\ln T$ [the inset in Fig.\ 6(a)], implying that the quantum critical fluctuations of the FM order, as suggested in Ca-doped SrRuO$_3$ \cite{rf:Kiyama98,rf:Huang2015}, are induced in a small number of unfrozen spins or clusters. 

In Fig.\ 6(b), we show the relation between $\gamma$ and $T^*$, obtained using $x$ as an implicit parameter. A clear $\gamma(T^*)-\gamma(T^*=0) \propto - T^*$ relation is found in the FM long-range order region for $x \le 0.3$, suggesting that the increase in $\gamma$ with increasing $x$ is mainly attributed to the recovery of the density of states at the Fermi level due to the suppression of the spin band splittings. In fact, a simple extrapolation to $T^* \to 0$ yields $\gamma \sim 65$ mJ/K$^2$ mol, which is close to the value of the isostructural nonmagnetic metal CaRuO$_3$ (73--82 mJ/K$^2$ mol)  \cite{rf:Cao97,rf:Kiyama98}. However, a significant deviation from the $\gamma(T^*)-\gamma(T^*=0) \propto - T^*$ relation due to an additional increase in $\gamma$ occurs in the cluster-glass region for $0.4 \le x \le 0.47$. This would reflect the entropy contributions associated with the clustering and the freezing of the spins; the very large broadening of the anomaly at $\sim T^*$ in $C_p/T$ may give rise to the quasi $T$-linear dependence of $C_p(T)$ at low temperatures. This feature has been observed in $C_p(T)$ of the typical spin-glass system AuFe \cite{rf:Wenger75}. In addition, it is likely that the emergence of the very weak quantum critical fluctuations also contributes to the enhancement of the $\gamma$ value. For all the La and K doping levels, the large $\gamma$ values (30--74 mJ/K$^2$ mol) are attributed to the Ru 4$d$ electronic contributions as their itinerant characteristics, as well as the freezing of their spins. In particular, the itinerant characteristics of Ru 4$d$ electrons are common in Ca-, La-, and La$_{0.5}$K$_{0.5}$-doped SrRuO$_3$ \cite{rf:Cao97,rf:Yoshimura97,rf:Kiyama98,rf:Kawasaki2014,rf:Kawasaki2016-2}.

\section{Discussion}
In Sr$_{1-x}$(La$_{0.5}$K$_{0.5}$)$_{x}$RuO$_3$, we observed the emergence of the cluster-glass state in the vicinity of the FM critical concentration: $x=0.5$. This feature is similar to the spontaneous phase separation in Sr$_{1-x}$Ca$_{x}$RuO$_3$ revealed by the $\mu$SR and magneto-optical experiments \cite{rf:Uemura2007,rf:Demko2012} and the cluster-glass formation in Sr$_{1-x}$La$_{x}$RuO$_3$ \cite{rf:Kawasaki2014,rf:Kawasaki2016-1}, but different from the non-Fermi-liquid state originating from the quantum critical fluctuations in Sr$_{1-x}$Ca$_{x}$RuO$_3$ suggested by the NMR and thermodynamic investigations \cite{rf:Yoshimura97,rf:Kiyama98,rf:Huang2015}. Furthermore, we found that Ru 4$d$ electrons have itinerant and localized characteristics simultaneously in Sr$_{1-x}$(La$_{0.5}$K$_{0.5}$)$_{x}$RuO$_3$. The former is realized by the large $\gamma$ value, and the latter is expected from the FM nano-cluster formation proposed by the Vogel-Fulcher analysis for $\chi'_{\rm ac}(T)$. We expect that the itinerant-localized dual nature involved in Ru 4$d$ electrons significantly affects the difference in the spin states around the FM critical region among the doped alloys. In this section, we suggest a possible origin of the avoided quantum criticality in Sr$_{1-x}$(La$_{0.5}$K$_{0.5}$)$_{x}$RuO$_3$ from the perspective of the dual nature of Ru 4$d$ electrons and ion doping effects.

\begin{figure}[tbp] 
\begin{center}
\includegraphics[viewport=0 0 462 459,keepaspectratio,width=0.42\textwidth]{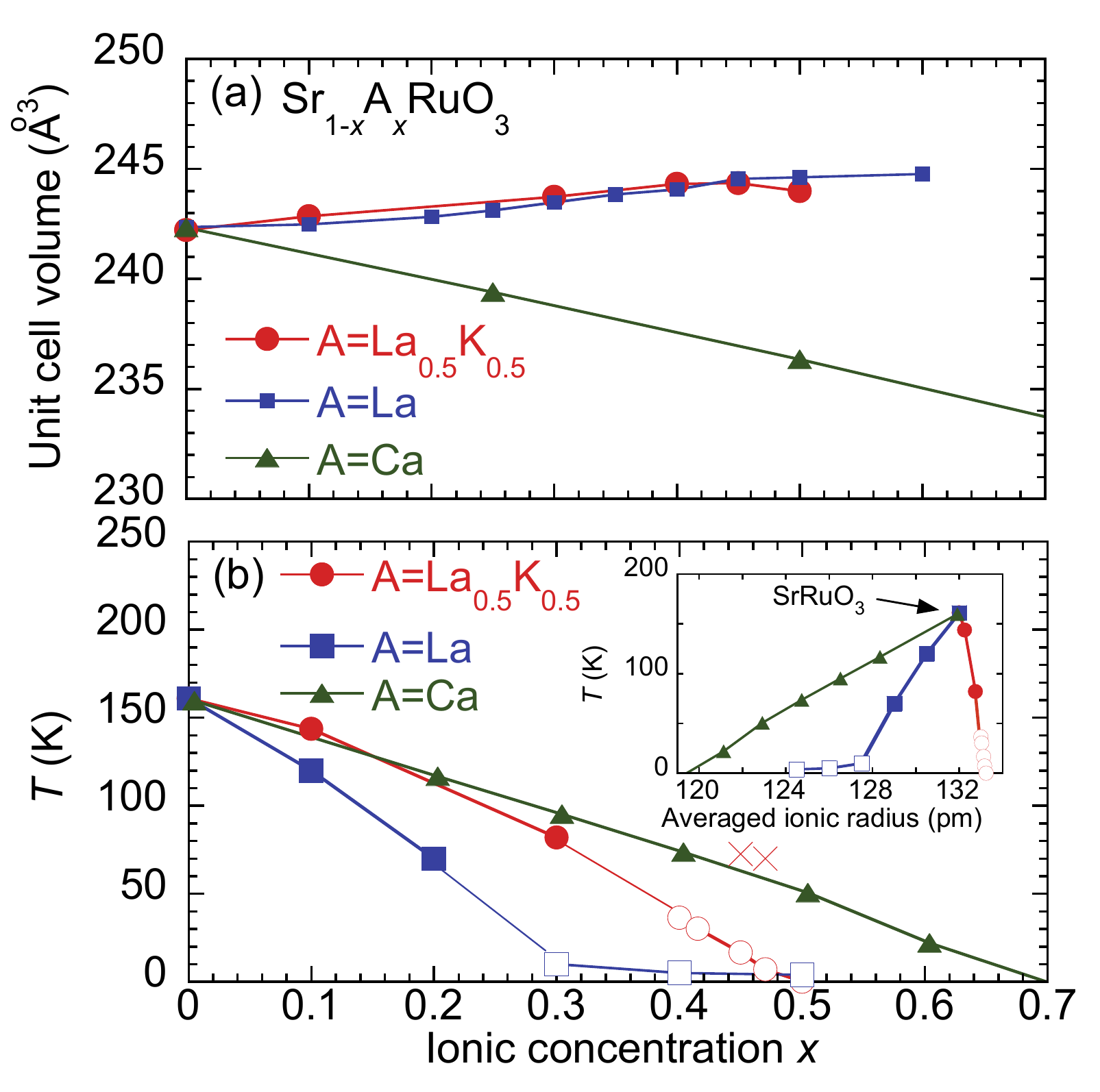}
\end{center}
\vspace{-12pt}
  \caption{
(a) The unit cell volume at room temperature and (b) the FM transition temperature plotted as a function of doped ionic concentration $x$ for Sr$_{1-x}$A$_{x}$RuO$_3$ with A=La$_{0.5}$K$_{0.5}$, La \cite{rf:Nakatsugawa2002,rf:Kawasaki2014}, and Ca \cite{rf:Cao97,rf:Yoshimura97}. The FM transition temperature is also plotted as a funcution of the averaged radius of the A-site ions in the inset of (b). In (b), the open and cross symbols indicate the freezing temperature of the cluster-glass state for La$_{0.5}$K$_{0.5}$- and La-doped SrRuO$_3$ and  $T_{\rm ir}$ for La$_{0.5}$K$_{0.5}$-doped SrRuO$_3$, respectively.
}
\end{figure}
First, it is natural to consider that the disorder effect induced by doping ions plays a vital role in the evolution of the FM clusters around the FM critical region. It has been argued that in Ca-doped SrRuO$_3$, the suppression of the FM order is coupled with the changes in the local ionic positions, such as ionic bond lengths and RuO$_6$ octahedra rotations in the distorted perovskite structure \cite{rf:Mazin97,rf:He2000,rf:Middey2011,rf:Cheng2013,rf:Wang2016,rf:Nguyen2020}. However, doping ions should generate the local disorder, possibly caused by the difference in the ionic radii between the doped ions (114 pm for Ca$^{2+}$, 117 pm for La$^{3+}$, and 152 pm for K$^+$) and Sr$^{2+}$ (132 pm). To roughly check the role of  disorder concerning the atomic positions in the evolution of the FM clusters, we compare the unit-cell volume among the doped alloys. Displayed in Fig.\ 7(a) are the ionic composition $x$ dependencies of the unit-cell volume for La$_{0.5}$K$_{0.5}$-, La-, and Ca-doped SrRuO$_3$ \cite{rf:Cao97, rf:Nakatsugawa2002}. Among the doped alloys, the largest volume change is generated in Ca-doped SrRuO$_3$, implying that disorder concerning the atomic positions would be largely induced around the doped Ca$^{2+}$ ions. However, this expectation does not seem compatible with the trend in the variations of the spin state among the alloys. The cluster-glass state is realized in La$_{0.5}$K$_{0.5}$- and La-doped SrRuO$_3$, whereas the quantum critical phenomena are often observed in Ca-doped SrRuO$_3$. A similar conclusion may be also derived from the relation between the FM transition temperature and the radii of the doped ions. For Ca-doped SrRuO$_3$, the reduction of the FM transition temperature by changing the average of the doped ionic radii is the smallest among the doped alloys [the inset in Fig.\ 7(b)]. This implies that the disorder concerning the atomic positions, accompanied by the variations in the average of the ionic radii, is not much relevant to the suppression of the FM order.

Instead, a key to understanding the origin of the discrepancy in the spin states at the FM critical region is derived from the features of the cluster-glass state found in La$_{0.5}$K$_{0.5}$- and La-doped SrRuO$_3$ \cite{rf:Kawasaki2014}. We have argued that in La$_{0.5}$K$_{0.5}$- and La-doped SrRuO$_3$, the suppression of $T^*$ is mainly due to the reduction in inter-cluster interactions rather than the shrinkage of the spin cluster in the FM critical regions, because $T_0$ is markedly reduced with $x$, while $E_{\rm a}$ is nearly independent of $x$. We expect that these features originate from the occurrence of local nano-sized Sr-rich regions in real space as a consequence of disorder concerning ionic-site occupations, possibly yielded by the large difference in the ionic radii between Sr$^{2+}$ and the doped ions regardless of the doped compounds. In such regions, the FM nano-droplets would be stable and pinned because itinerant Ru 4$d$ electrons have a very small mean-free path as a characteristic of the bad metal. As the doping level is increased, the distance between the Sr-rich regions becomes large, while the lower limit of the Sr-rich region size for stabilizing the FM droplet is unchanged, leading to the reduction in $T_0$ and the nearly unchanged $E_{\rm a}$ value, as revealed by the present $\chi_{\rm ac}'(T)$ experiments. If this is the case, whether the cluster-glass state or quantum critical state evolves should depend significantly on the sample preparations and conditions, as observed in Sr$_{1-x}$Ca$_{x}$RuO$_3$ \cite{rf:Yoshimura97,rf:Kiyama98,rf:Huang2015,rf:Uemura2007,rf:Demko2012}. The interpretation above further proposes that quantum critical behavior is substantially governed by the so-called Griffiths instability \cite{rf:Vojta2010} even in a disorder-reduced sample of Sr$_{1-x}$Ca$_{x}$RuO$_3$. This possibility has also been inferred from the anomalously small dynamic critical exponent concerning FM quantum critical fluctuations \cite{rf:Huang2015}. 

The coupling between the correlated lattice disorder and the FM cluster formation has been suggested from the observation of the smeared quantum phase transition in the epitaxial film of Sr$_{1-x}$Ca$_{x}$RuO$_3$ \cite{rf:Demko2012} and the comparison between Ca- and Ba-doped alloys \cite{rf:Cheng2013}. In addition, the recent structural analyses and the magnetic susceptibility measurement for Sr$_{1-x}$Ca$_{x}$RuO$_3$ revealed that the inhomogeneity of the Ca/Sr distribution can be significantly reduced by tuning the sample preparation condition \cite{rf:Nguyen2020}. In general, it is hard to detect such local heterogeneity with conventional x-ray diffraction techniques. Thus, we expect that investigations of the precise local crystal structure using atomic probes would provide comprehensive understanding regarding this local disorder effect on the magnetic ground state around the FM critical region.

Second, we argue a possible carrier doping effect on the suppression of the FM order. Figure 7(b) shows the FM transition temperature and the freezing temperature plotted as a function of ionic composition $x$ for La$_{0.5}$K$_{0.5}\textrm{-}$, La-, and Ca-doped SrRuO$_3$ \cite{rf:Yoshimura97,rf:Nakatsugawa2002}. For $x \le 0.3$, the decreasing rates of the FM transition temperature by $x$ in La$_{0.5}$K$_{0.5}$- and Ca-doped SrRuO$_3$ are comparable, but are clearly smaller than that in La-doped SrRuO$_3$. This difference could be simply ascribed to the effect of carrier doping, because it is expected that the La$^{3+}$ ion acts as an electron dopant in SrRuO$_3$, whereas the La$^{3+}_{0.5}$K$^{+}_{0.5}$ and Ca$^{2+}$ ions do not. For $x \ge 0.4$, however, $T^*$ of the La$_{0.5}$K$_{0.5}$-doped alloys decreases more rapidly than the FM transition temperature of the Ca-doped alloys although the doped ions have nominally the same valence as Sr$^{2+}$ in both alloys. The FM order is replaced by the cluster-glass state at this $x$ range in La$_{0.5}$K$_{0.5}$-doped SrRuO$_3$. Thus, we consider that the rapid decrease in $T^*$ could be caused by local heterogeneity of the Sr ion distribution rather than the nominal carrier doping effect, as argued above. In this regard, we further suggest that the onset temperature of the FM cluster formation in La$_{0.5}$K$_{0.5}$-doped alloys, which is expected to nearly correspond to $T_{\rm ir}$ [the cross symbols in Fig.\ 7(b)], would be comparable to the FM transition temperature of the Ca-doped alloys for $x\ge 0.4$. Such an FM cluster formation for $T>T^*$ was observed with the $\mu$SR experiment for La-doped SrRuO$_3$ \cite{rf:Kawasaki2016-1}. Despite these considerations, the local charge distribution around Sr$^{2+}$ and the doped ions, as a consequence of the charge compensation by the co-doping of the La$^{3+}$ and K$^{+}$ ions, is still unclear in Sr$_{1-x}$(La$_{0.5}$K$_{0.5}$)$_{x}$RuO$_3$ at present. It has been suggested that in La$_{0.5}$Na$_{0.5}$-doped SrRuO$_3$, inhomogeneous charge compensation may give rise to a local disorder effect around the doped ions, and it then reduces the FM order more rapidly in the intermediate La$_{0.5}$Na$_{0.5}$-doping level, although it is still unclear whether such an effect also becomes an origin of the FM nano-cluster formation \cite{rf:He2000}. Photoemission spectroscopy measurements are expected to provide details of the electronic state in doped alloys.

\section{Conclusion}
In Sr$_{1-x}$(La$_{0.5}$K$_{0.5}$)$_{x}$RuO$_3$, the FM order originating from itinerant Ru 4$d$ electrons is suppressed by increasing the concentration of doped La and K ions, and then disappears at $x=0.5$ \cite{rf:Shuba2006}. The present investigation for Sr$_{1-x}$(La$_{0.5}$K$_{0.5}$)$_{x}$RuO$_3$ using dc magnetization, ac magnetic susceptibility, and specific heat experiments revealed that the FM quantum phase transition is avoided, and it is replaced by cluster-glass formation at the FM critical region of $0.4 \le x \le 0.47$. Suppression of FM quantum critical fluctuations was shown by non-diverging behavior in $C_p/T$ for $T\to 0$. Alternatively, we suggested the emergence of the cluster-glass state from the Vogel-Fulcher analysis for the frequency variations in $\chi_{\rm ac}'(T)$ and the observation of increased $\gamma$ values although there was no salient anomaly at $T^*$ in $C_p/T$. We discussed the origin of the avoided quantum phase transition by comparing the features of FM suppression in Sr$_{1-x}$(La$_{0.5}$K$_{0.5}$)$_{x}$RuO$_3$ with those in Ca- and La-doped SrRuO$_3$, and suggested that a local correlated disorder effect and the very small coherence of itinerant Ru 4$d$ electrons are responsible for the cluster-glass formation instead of the quantum phase transition.

\begin{acknowledgments}
We are grateful to A. Kondo for his experimental support. M.Y. expresses gratitude to T. Nakano and I. Kawasaki for fruitful discussions. This research was partly carried out as joint research in ISSP, and was supported by JSPS KAKENHI Grant Nos. 17K05529 and 20K03852.
\end{acknowledgments}


\end{document}